\documentstyle[11pt,paspconf,164page,164def,psfig]{article}
\begin{document}

\title{A MERLIN and VLBI Survey of Faint Compact Radio Sources}

\index{Surveys} 
\index{FIRST} 
\index{MERLIN} 

\markboth{Garrett \& Garrington}{A MERLIN Survey of Faint Compact Radio 
Sources from the VLA FIRST Survey}

\author{M.A.~Garrett}
\affil{JIVE, Postbus 2, 7990~AA, Dwingeloo, NL.}

\author{S.T.~Garrington} 
\affil{U. of Manchester, NRAL, Jodrell Bank, Macclesfield, Cheshire, SK11 9DL, UK.} 

\begin{abstract}  
  We have selected a field from the VLA FIRST Survey which is typical
  in all aspects except one: it contains a bright, but extremely
  compact VLBI calibrator - J1159+291.  127 unresolved FIRST sources
  with $S_{\rm FIRST}\ge 10$~mJy lie within $2.5^{\circ}$ of this
  calibrator. $\lambda 6$~cm MERLIN observations with a resolution
  $\sim 60$~mas detect around half the sources.  These detections
  form the basis of a sample of faint but compact radio sources which
  are ideally suited to follow-up VLBI observations.
\end{abstract} 

\section{The Faint Source Sample and Observing Strategy}

Present day, large-scale VLBI surveys have so far focussed on samples
of relatively bright (and often flat spectrum) radio sources with
total flux densities typically $> 350$~mJy.  By adopting
these selection criteria, VLBI surveys have been remarkably
successful, but by only targeting the very brightest objects, our
current knowledge and general understanding of compact radio sources
is surely incomplete.

The technique of phase-referencing now allows sources as faint as
1~mJy per beam to be detected and reliably imaged. We have recently 
embarked on a faint source VLBI survey, which employs a novel
observing strategy. We have defined a sample of 127 sources which were
drawn from the VLA FIRST survey (Becker, R.~H., White, R.~L., \&
Helfand, D.~J. 1995) and these satisfy the following simple selection
criteria, all sources: (i) lie within a ``selection box''
$3^{\circ}\times4^{\circ}$ wide, centred on the VLBI calibrator J1159+291,
(ii) have measured FIRST sizes $ \leq 5 $~arcsec and,
(iii) have FIRST peak fluxes $> 10$~mJy/beam.

The advantages of this approach are considerable: (i) a relatively
unbiased source sample arises naturally, (ii) by focusing on one small
field scheduling is effortless and observing efficiency maximized (the
long telescope slew times usually associated with snapshot surveys are
eliminated), (iii) measurement of source redshifts (using multi-fiber
optical spectroscopy) requires only a modest amount of 
telescope time, (iv) by careful selection of a ``model'' calibrator,
the data analysis can be largely automated and high SNR (delay, rate
and phase) solutions can be obtained with very short calibrator
integration times, (v) the offset on the sky between the source and
the calibrator is typically much less than $2.5^{\circ}$, even for
relatively large samples, and finally (vi) the inverse
process of attempting to identify phase-calibrators for every target  
in a large faint source sample, is happily avoided.  

\section{MERLIN Observations} 

\begin{figure}[htb]
\vspace{45mm}
\begin{picture}(120,35)
\put(-20,170){\includegraphics{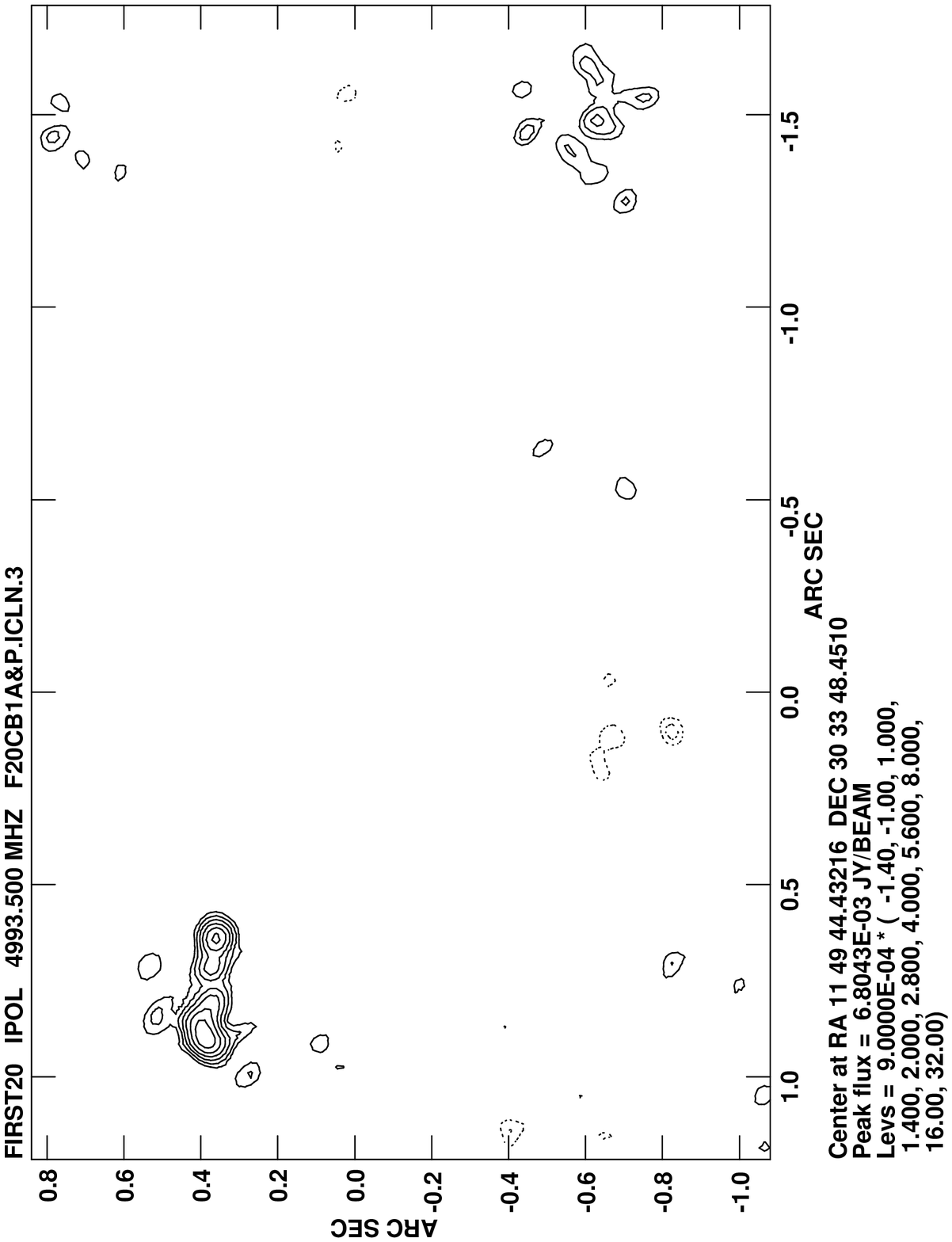}}
\put(203,-45){\includegraphics{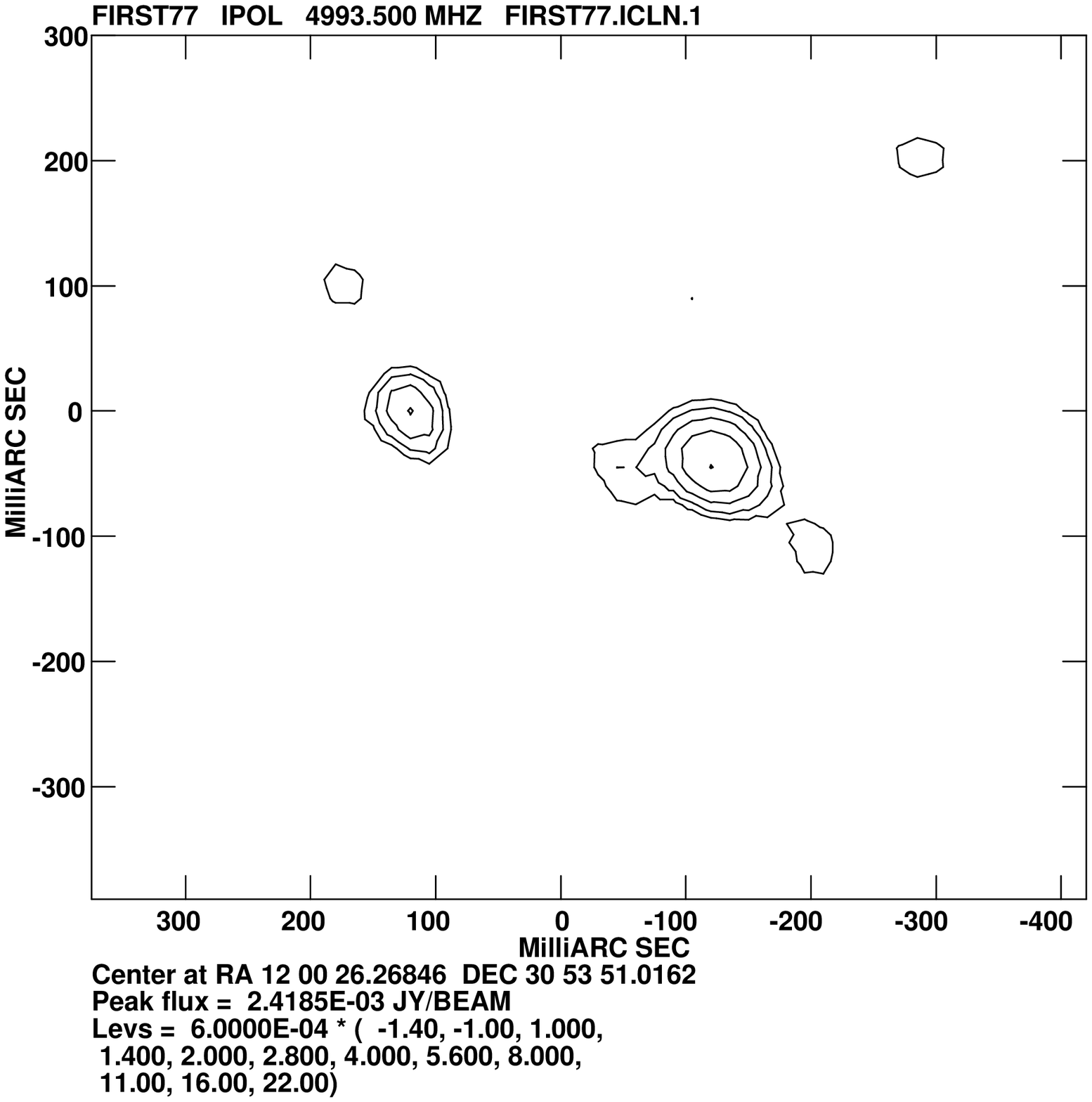}}
\end{picture}
\caption{Two of the faint FIRST sources which were detected by
MERLIN at 6cm.}
\label{fig1}   
\end{figure}

Our 127 FIRST sources were observed by MERLIN at $\lambda 6$~cm in
January this year. Each source was observed in snapshot,
phase-reference mode for a total of 15 minutes spread over 12~hr in
hour angle. The data were analysed automatically with the phase and
amplitude corrections of the calibrator (J1159+291) being applied to
each target source. The final maps had an r.m.s.  noise of
0.3~mJy/beam and a resolution of 60~mas (see Fig. \ref{fig1}). Of the 127
sources we observed with MERLIN, 68 were clearly detected ($7
\sigma$).  5 of these were observed to be very extended and are most
likely the lobes of extended radio sources which appear as ``doubles''
in the FIRST maps. The MERLIN detection rate is impressive, especially
when one considers that the $\lambda6$~cm beam area is a factor of 10000
smaller than FIRST's.

Preliminary MERLIN $\lambda 18$~cm images show that the majority of
sources have little intermediate structure (0.5 -- 3.0 arcsec).  Our
high detection rate is interesting, the evolutionary
models of Wall \& Jackson 1997 predict that the fraction of
flat spectrum objects should decrease to 10-20\% at these flux levels.

\section{Motivation and Current Status}

For the first time, the combination of the FIRST survey, MERLIN filter
observations and the phase-reference technique, allows us to reliably
image and systematically classify a large number of faint, compact
radio sources which are 1 to 2 orders of magnitude fainter than those
targeted in previous VLBI surveys. We intend to compare the properties
of these faint sources with their brighter cousins. By going deeper we
hope to uncover a few surprises too.

Phase-reference MERLIN-VLBI observations have already begun at
$\lambda 18$cm and further $\lambda6$cm VLBI and MERLIN observations
are planned.  In the spirit of the FIRST project the MERLIN
$\lambda\lambda 6$ and 18~cm images and further information are
available on-line at: {\tt http://www.nfra.nl/\verb+~+mag/first.html}.

\end{document}